%% file: main.tex
\documentclass[11pt]{article}
\usepackage{mathrsfs}
\usepackage{amsmath}
\usepackage[colorlinks=true, allcolors=blue, urlcolor=black]{hyperref}

\makeatletter
\newcommand{\removelatexerror}{\let\@latex@error\@gobble}
\makeatother
\usepackage{amsmath}
\usepackage{amsthm}
\usepackage{amsfonts}
\usepackage{amssymb}
\usepackage{graphicx}
\usepackage{dsfont}
\usepackage{csquotes}
\usepackage{mathtools}
\usepackage{cleveref}
\usepackage{pgfplots}
\pgfplotsset{compat=1.18}
\usepackage{bbm}
\usepackage{stmaryrd}

\usepackage{complexity}
\usepackage{enumitem}
\usepackage{tikz}
\usetikzlibrary{positioning}
\usepackage{thm-restate}
\usepackage{float}
\usetikzlibrary{calc}
\usepackage{thm-restate}
\usetikzlibrary{shapes,decorations.markings}
\usepackage{caption}
\usepackage{subcaption}
\usepackage{comment}

\usepackage{lipsum}

\newtheorem{theorem}{Theorem}[section]
\newtheorem{observation}[theorem]{Observation}

\newtheorem{claim}[theorem]{Claim}
\newtheorem{lemma}[theorem]{Lemma}
\newtheorem{proposition}[theorem]{Proposition}
\newtheorem{question}[theorem]{Question}
\newtheorem{openquestion}{Open Question}
\newtheorem{remark}[theorem]{Remark}

\newtheorem{corollary}[theorem]{Corollary}

\renewcommand{\E}{\mathbb{E}}

\newcommand{\MAJ}{\operatorname{MAJ}}

\newcommand{\Inf}[0]{\operatorname{Inf}}

\renewcommand{\R}[0]{\mathsf{R}}
\newcommand{\switch}{\operatorname{sw}}

\usepackage[english]{babel}
\usepackage[T1]{fontenc}
\usepackage{caption}

\usepackage[ruled,vlined,linesnumbered]{algorithm2e}
\crefname{algocf}{Algorithm}{Algorithms}

\usepackage[margin=1in]{geometry}

\title{Weak Composition Theorems for Randomized Communication}
\author{Pachara Sawettamalya\footnote{Princeton University. \ \href{mailto:ps3122@princeton.edu}{\url{ps3122@princeton.edu}}}}
\date{\today}

\usepackage[backend=biber, style=alphabetic, doi=false, url=false, isbn=false, minalphanames = 3, maxalphanames = 4, maxbibnames=10]{biblatex}
\begin{document}
\maketitle

\vspace{-0.77cm}

\input{0-abstract}

\thispagestyle{empty}
\pagenumbering{roman}
\pagenumbering{arabic}

\input{1-intro}

\input{2-prelim}

\input{3-approach}

\input{4-lifting}

\input{5-symmetric}

\section{Acknowledgement}
The author thanks Huacheng Yu for many insightful discussions, as well as the anonymous ITCS 2027 reviewers for their helpful comments.

\paragraph{AI Usage.} We used ChatGPT-5.6 to assist with language editing. All mathematical content, including the literature review, statements, and proofs, was developed by the author without machine assistance.
\printbibliography

\appendix 

\input{A-even_majority}

\end{document}

%% file: 0-abstract.tex
\begin{abstract}

We present two results towards the randomized communication complexity of composed functions. Our contributions include:

\begin{itemize}
    \item \textbf{Lifting-by-Influence.} For any binary function $f$ and any outer gadget $g:\{0,1\}^n \rightarrow \mathbb{R}$, we show that up to additive terms, computing $g \circ f^n$ information-theoretically requires solving at least $\operatorname{Inf}(g)$ instances of $f$, where $\operatorname{Inf}(g)$ denotes the total influence of $g$. To our knowledge, this is the first composition lower bound for randomized communication that applies to \emph{every} choice of $f$ and $g$.

\item \textbf{Symmetric Composition.}  When $g$ is a symmetric function with a value transition at Hamming weight $d$, we strengthen the number of required instances to $\min\{d,n+1-d\}$. As an implication, this recovers the previously known direct-sum theorem, XOR lemma, and majority lemma.
\end{itemize}

Our proofs develop novel and distinctive usages of the classical \emph{input embedding} argument of Barak, Braverman, Chen, and Rao (SICOMP 2013). As an additional ingredient, we introduce a new public sampling framework that avoids auxiliary communication during the recovery phase of the protocol, ultimately reducing the extra cost from $O(n)$ bits of information to just $O(1)$.




\end{abstract}

%% file: 1-intro.tex
\vspace{-0.5cm}

\section{Introduction}

One of the central questions in computational complexity is to understand how the hardness of a function behaves under composition. That is, given a function $f$, one can construct a new function by combining multiple instances of $f$ through another ``outer gadget'' $g$. But how much harder will this task become? This motivates the study of the complexity of composed functions.

\begin{question}[Composition theorem]
Let $f$ be a binary function and $g:\{0,1\}^n\to\mathbb{R}$ be an outer gadget. How does the complexity of $g\circ f^n$ relate to the complexities of $f$ and $g$?
\label{ques:composition}
\end{question}

From the upper bound perspective, a straightforward solution is to solve all $n$ instances of $f$ and then compile the outputs through $g$. This approach uses $n$ times the cost required to compute $f$ (i.e. the \emph{scaling} is $n$). In fact, one might improve this scaling to the \emph{decision tree complexity} of $g$ by  only evaluating instances of $f$ as guided by such tree. The main challenge is yet to understand the lower bound direction: at what cost computing $g\circ f^n$ require in comparison to computing $f$? There is, however, no universal answer to this question as it may depend delicately on the descriptions of $f$ and $g$, the cost being measured, and the computational model under consideration.

In this note, we study such question in the standard two-party randomized communication setting \cite{Yao79}, where Alice receives an input $x$, Bob receives an input $y$, and together they seek to compute a joint function $f(x,y)$ while exchanging as little communication as possible. The communication complexity $\R(f)$ is the minimum number of bits needed to compute $f$ with an error bound of $1/3$. We also consider the \emph{information} complexity, denoted $\IC_{\mu}(f)$, which measures the \emph{information} revealed about the inputs over a distribution $\mu$.\footnote{For introductory purposes, we defer the formal definition of these quantities to \Cref{sec:prelim}.} The same set of notions also apply for the composed function $g\circ f^n$ where Alice receives  $(x_1,\ldots,x_n)$ and Bob receives $(y_1,\ldots,y_n)$ as their inputs, and together they wish to determine $g\left(f(x_1,y_1),\ldots,f(x_n,y_n)\right)$.


Our first result establishes a generic lower bound for the communication complexity of composed functions. We show that information-theoretically, computing $g\circ f^n$ requires solving roughly $\Inf(g)$ instances of $f$, where $\Inf(g)$ denotes the \emph{total influence} of $g$. While previous works study such composition lower bounds for specific inner functions $f$ \cite{RazM99, GoosP018, HatamiHL18,LoffM19,GoosPW20, ChattopadhyayFK21} or specific outer functions $g$ \cite{BBCR13, Yu22, IR24b, SawettamalyaY25, SawettamalyaY26}, our result gives, to the best of our knowledge, the first lower bound that applies to \emph{every} choice of $f$ and $g$.



\begin{theorem}[Lifting-by-influence]
\label{thm:lifting_intro}
For every function $f:\mathcal{X}\times \mathcal{Y} \rightarrow \{0,1\}$, gadget $g:\{0,1\}^n \rightarrow \mathbb{R}$, and input distribution $\mu$ with $\E_{\mu}[f(x,y)] = \frac{1}{2}$, we have 
\begin{align*}
\R(g\circ f^n)\geq \Omega\left(\Inf(g)\cdot \IC_\mu(f)-n\right).
\end{align*}
\end{theorem}


This theorem recovers several previously known results. When $g$ is the identity function (total influence $n$), the composition corresponds to computing $n$ instances of $f$, placing us in the realm of the \emph{direct-sum} problem. The seminal work of Barak, Braverman, Chen, and Rao \cite{BBCR13} established a direct-sum property for information complexity and a weak direct-sum property for communication complexity. Their proof formalized the \emph{input embedding} argument which has since become a central tool in deriving direct-sum-type results.\footnote{Historically, although the approach was first made explicit by \cite{BBCR13}, its origins can be traced back to earlier works; see, for example, \cite{Razborov92,CSWY01,BYJKS04}. We will recall the embedding argument in \Cref{subsec:BBCR}.} For instance, it has inspired \emph{direct-product} theorems, showing that the success probability of computing all $n$ instances of $f$ decays at exponential rates \cite{BRWY13,NelsonY19}. Another natural gadget that has received much attention is the parity function $\mathrm{XOR}(z_1,\ldots,z_n):=z_1\oplus\cdots\oplus z_n$, which also has total influence $n$. This leads to the line of work on the \emph{XOR lemmas}, where several results establish optimal lower bounds with an $\Omega(n)$ scaling \cite{BBCR13,Yu22,IR24b,SawettamalyaY25}.  These, as well, can be recovered via \Cref{thm:lifting_intro}.


For more than a decade, identity and parity have remained the only gadgets for which the complexity of composition was well understood. While \Cref{thm:lifting_intro} provides a starting baseline for arbitrary gadgets, the total influence can be a loose structural parameter when applied to specific choices of $g$. A prime example is the \emph{majority} function $\MAJ(z_1,\ldots,z_n):=\mathbbm{1}(z_1+\cdots+z_n\ge n/2)$, where it is natural to conjecture that computing $\MAJ\circ f^n$ requires essentially solving all $n$ instances of $f$, i.e. the scaling is $\Omega(n)$. Indeed, an analogous \emph{majority lemma} has been known for the randomized query complexity \cite{Ben-DavidGK020,GoosM21}, yet its communication analogue has been missing until very recently. One possible explanation is that the embedding techniques frequented by the direct-sum theorems and XOR lemmas do not readily extend to majority composition where each individual answer of $f$ is too insensitive to the overall majority answer. This barrier is also reflected in our first result. That is, applying \Cref{thm:lifting_intro} with $g=\MAJ$ yields only a multiplicative factor $\Inf(\MAJ)=\Theta(\sqrt{n})$ which, although non-trivial, falls significantly short of $\Omega(n)$.


It was only in the recent work of Sawettamalya and Yu \cite{SawettamalyaY26} that the majority lemma for randomized communication was finally established with the conjectured $\Omega(n)$ scaling. Their proof bypasses the said barrier via an indirect approach that combines sophisticated variants of the embedding argument with other technical ingredients lifted from prior works.\footnote{Indeed, \cite{SawettamalyaY26} proves the stronger \emph{gap-majority lemma}, from which the majority lemma arises as a corollary.} To this, our second result yields a new and succinct proof of the majority lemma. Conceptually, we show that the classical embedding approach can, in fact, be adapted specifically to majority composition in a simple, direct, and self-contained manner.

Crucially, we stress that our second result extends beyond the majority lemma. We obtain a composition theorem that is applicable for \emph{every symmetric} outer function $g$, where the scaling factor is dictated by the location of the transitions of $g$.

\begin{theorem}[Symmetric composition theorem]
\label{thm:sym_intro}
Let $g:\{0,1\}^n \rightarrow \mathbb{R}$ be a non-constant symmetric function, and let $d\in[n]$ be such that the evaluations of $g$ at Hamming weights $d-1$ and $d$ differ. For every function $f:\mathcal{X}\times\mathcal{Y}\rightarrow\{0,1\}$ and every input distribution $\mu$, we have
\begin{align*}
\R(g\circ f^n)\ge \Omega\left(\min\{d,n+1-d\}\right)\cdot\bigl(\IC_\mu(f)-O(1)\bigr).
\end{align*}
\end{theorem}

We highlight two consequences of \Cref{thm:sym_intro}. First, like \Cref{thm:lifting_intro}, it also unifies several previously known composition results. Since every Hamming weight is a transition point of the parity function, choosing any $d=\Theta(n)$ recovers the XOR lemma, which then implies the direct-sum theorem. It can also recover the majority lemma (which \emph{cannot} be obtained via \Cref{thm:lifting_intro}) by setting $d=\frac{n+1}{2}$ when $n$ is odd. When $n$ is even, however, the argument no longer applies because the tie-breaking at Hamming weight $n/2$ may be arbitrary. Nevertheless, a simple reduction recovers the same $\Omega(n)$ lower bound. We defer this proof to Appendix \ref{appendix:even_maj}.

Second, \Cref{thm:sym_intro} makes progress toward characterizing ``hard'' gadgets. An open question in this domain is to identify explicit gadgets $g$ for which computing $g\circ f^n$ asymptotically requires solving all $n$ instances of $f$. Prior to our work, such statement was known only for the identity, parity, majority, and (gap-)majority functions. Our theorem substantially broadens this picture by showing that every symmetric function with a transition at $\Theta(n)$ also admits this property.

\subsection{From Information to Communication}

A limitation of \Cref{thm:lifting_intro,thm:sym_intro} is that they lower bound the communication complexity of $g\circ f^n$ by the \emph{information} complexity of $f$. Although these two  measures oftentimes asymptotically coincide for many canonical communication problems, they are not generally interchangeable  and may even differ exponentially \cite{GanorKR14,GanorKR16}.

Regardless, our information-theoretic lower bounds can be converted into worst-case communication lower bounds at the cost of a square-root loss in the scaling. The key ingredient is the \emph{internal compression} scheme of \cite{BBCR13} which shows that any protocol with communication cost $C$ and information cost $I$ can be simulated using roughly $\sqrt{C\cdot I}$ bits of communication. This routine was the main ingredient behind the breakthrough direct-sum theorem of \cite{BBCR13}. Applying the same argument to \Cref{thm:lifting_intro,thm:sym_intro} yields the following corollary (proof omitted). To our knowledge, this is also the first generic composition theorem for worst-case communication, despite its seemingly weak bounds on the scaling factor.


\begin{corollary}
Suppose that the function $f:\mathcal{X}\times\mathcal{Y}\to\{0,1\}$ and the gadget $g:\{0,1\}^n\to\mathbb{R}$ satisfy $\R(g\circ f^n)\ge \Omega(n)$. Ignoring polylogarithmic factors, computing $g\circ f^n$ requires $\sqrt{\Inf(g)}$ times the amount of \emph{communication} for computing $f$. If $g$ is symmetric, the scaling improves to $\sqrt{\switch(g)}$ where $\switch(g)$ denotes the maximum value of $\min\{d,n+1-d\}$ over all transitions $d$ of $g$.
\label{cor}
\end{corollary}


\subsection{Related Open Questions}


An important open question in communication complexity is to understand the full scope of \emph{query-to-communication} lifting theorems: can the communication complexity of a composed function $g \circ f^n$ be characterized by the \emph{query} complexity of the outer function $g$? This is known to hold for several specific inner functions $f$, most notably indexing and inner product \cite{RazM99,Sherstov11,GoosP018,HatamiHL18,LoffM19,GoosPW20,ChattopadhyayFK21}. Whether this phenomenon holds in generality remains a major open problem. We reiterate the information-theoretic variant of this question, posted by \cite{ChattopadhyayFK21}.

\begin{openquestion}
    Prove or disprove: up to lower-order terms, computing $g \circ f^n$ requires $\mathsf{DT}(g)$ times the \emph{information} required to compute $f$ where $\mathsf{DT}(g)$ denotes the decision tree complexity of $g$.
\end{openquestion}

One might also consider a weaker version of the problem by asking for lower bounds that depend only \emph{polynomially} on the information complexity of $f$ and $\mathsf{DT}(g)$. While such bounds are known in deterministic communication \cite{Iyer25}, no analogous result is known for the randomized setting.

The majority lemma also suggests a related question. To compute $\MAJ \circ f^n$ with constant error, one may compute each copy of $f$ with reduced error $O(1/n)$ and apply a union bound. Our proof does not account for such error reduction, and it is natural to ask whether such feature is needed. 

\begin{openquestion}
Prove or disprove: up to additive terms, computing $\MAJ \circ f^n$ with error $1/3$ requires $\Omega(n)$ times the communication (or information) required to compute $f$ with error $n^{-0.01}$.
\end{openquestion}

We conjecture the answer to be positive as an analogous result is known for randomized query complexity \cite{GoosM21}. In parallel, the error reduction is shown to be necessary for both the identity and XOR gadgets in the communication setting \cite{SawettamalyaY25}.

%% file: 2-prelim.tex
\section{Preliminaries}
\label{sec:prelim}

For any natural number $n$ and $a\leq b$, let $[n]$ denote $\{1,\ldots,n\}$ and let $[a,b]$ denote $\{a,\ldots,b\}$.
For a function $f: \mathcal{X} \times \mathcal{Y} \rightarrow \{0,1\}$ and an outer gadget $g: \{0,1\}^n \rightarrow \mathbb{R}$, define $g \circ f^n$ as
\begin{align*}
    (g \circ f^n)(X_1,\ldots,X_n,Y_1,\ldots,Y_n) := g\left(f(X_1,Y_1),\ldots,f(X_n,Y_n)\right).
\end{align*}
For each $i \in [n]$, we write $Z_i = f(X_i,Y_i)$. For a set $S \subseteq [n]$, let $X_S$ denote $\{X_s\}_{s\in S}$. We may shorthand $X_{<i}$ for $(X_1,\ldots,X_{i-1})$, $X_{>i}$ for $(X_{i+1},\ldots,X_n)$, $\overline{X}$ for $(X_1,\ldots,X_n)$, and $X_{-i}$ for $X_{[n]\setminus\{i\}}$. 

\paragraph{Symmetric functions.}
A function $g:\{0,1\}^n\rightarrow\mathbb{R}$ is \emph{symmetric} if its evaluation at $z\in\{0,1\}^n$ depends only on the Hamming weight of $z$, i.e., there exists $\hat{g}:[0,n] \rightarrow\mathbb{R}$ such that $g(z)=\hat{g}(\|z\|_1)$. We say that $d\in[n]$ is a \emph{transition} of $g$ iff $\hat{g}(d-1)\neq\hat{g}(d)$. Among all transitions, we are particularly interested in the most middling one. We quantify this by
\begin{align*}
\switch(g):=
\max_{d\in[n]\text{ s.t. }\hat{g}(d-1)\neq\hat{g}(d)}
\ \min\{d,n+1-d\}.
\end{align*}


\paragraph{Influence.} For a function $g:\{0,1\}^n \rightarrow \mathbb{R}$ over $n$ variables $(z_1,\ldots,z_n)$, the \emph{influence} of $z_i$ is the probability that flipping the values of $z_i$ changes the value of the function. This is denoted by  
\begin{align*}
    \Inf_i(g) := \Pr_{z \sim \{0,1\}^n}\Big[g(z) \ne g(z \oplus e_i)\Big].
\end{align*}
The \emph{total influence} of $g$ is denoted by $\Inf(g) := \sum_{i=1}^{n} \Inf_i(g).$

\paragraph{Mutual information.} Let $X,Y,Z$ be random variables in a probability space $\pi$. The \emph{mutual information} between $X$ and $Y$ conditioned on $Z$ is defined by \begin{align*}
    I_{\pi}(X:Y \mid Z) := H(X \mid Z) - H(X \mid YZ)
\end{align*} where $H(\cdot)$ denotes the Shannon entropy. We will utilize these facts about mutual information.
\begin{proposition}[Chain rule] $I_{\pi}(XY:Z) = I_{\pi}(X:Z) + I_{\pi}(Y:Z \mid X)$.
\end{proposition}


\begin{lemma} Let $W$ be an event that occurs with probability $p$ in the probability space $\pi$. Then, for any random variables $X,Y,Z$, we have $I_{\pi \mid W}(X:Y \mid Z) \leq \frac{1}{p} \cdot \left( I_{\pi}(X:Y \mid Z) + 1\right)$.
\label{lem:helper}
\end{lemma}

\begin{proof} Let $\mathbbm{1}_W$ denote an indicator random variable of event $W$. Observe that $I_{\pi}(X:Y \mid Z\mathbbm{1}_W) \geq p \cdot I_{\pi \mid W}(X:Y \mid Z).$ By the chain rule of mutual information, we have 
\begin{align*}
    I_{\pi}(X:Y \mid Z\mathbbm{1}_W) = I_{\pi}(X\mathbbm{1}_W:Y \mid Z) - I_{\pi}(\mathbbm{1}_W:Y \mid Z) \leq I_{\pi}(X\mathbbm{1}_W:Y \mid Z)
    \end{align*}
and
\begin{align*} I_{\pi}(X\mathbbm{1}_W:Y \mid Z) = I_{\pi}(X:Y \mid Z) + I_{\pi}(\mathbbm{1}_W:Y \mid ZX) \leq I_{\pi}(X:Y \mid Z) + 1.\end{align*}
where the final inequality follows $ I_{\pi}(\mathbbm{1}_W:Y \mid ZX) \leq H(\mathbbm{1}_W) \leq 1$. Combining all inequalities conclude the proof.
\end{proof}

\paragraph{Communication protocols.} A task of computing a function $f:\mathcal{X} \times \mathcal{Y} \rightarrow \{0,1\}$ in two-party communication is as follows. Alice receives $X \in \mathcal{X}$ and Bob receives $Y \in \mathcal{Y}$. They may take turns sending messages where each message is generated from the past messages, the sender's own input, and the public randomness. Towards the end, they come up with their belief of $f(X,Y)$. The scheme of which players generate messages and the answer are called \emph{communication protocols}. 

We also adopt a \emph{distributional} view of the protocol. That is, suppose that the input pairs $(X,Y)$ is drawn from a specified distribution $\mu$. We interpret the public randomness by a random string $R$. The sequence of messages $M$ is then obtained via executing the protocol on the inputs $(X,Y)$ using randomness $R$. Thus, the protocol can be viewed as a joint distribution of $(X,Y,M,R)$.

The notions also remain the same for the composed function $g \circ f^n$, barring that the inputs $\overline{X} = (X_1,\ldots,X_n) \in \mathcal{X}^n$ are given to Alice and $\overline{Y} = (Y_1,\ldots,Y_n) \in \mathcal{Y}^n$ are given to Bob.

\paragraph{Costs.}  For a protocol $\pi = (X,Y,M,R)$, its \emph{communication cost}, denoted $\mathsf{CC}(\pi)$, is the worst-case message length $|M|$. Following the notations of \cite{BBCR13}, the \emph{information cost} of $\pi$ is defined as  $$\IC(\pi) := I_{\pi}(M:X \mid YR) + I_{\pi}(M:Y \mid XR).$$ 
Intuitively, the information cost measures the amount of information that players gain from $\pi$. To see this, consider Bob's perspective of the protocol. Initially, he only knows $Y$ and $R$. At the end of $\pi$, he learns $M$. Thus, Bob gains $I_{\pi}(M:X\mid YR)$ bits of information about Alice's input $X$.

\paragraph{Communication and Information Complexity.} For a function $f$, the \emph{randomized communication complexity} of $f$, denoted $\R(f)$, is the minimum communication cost of a protocol that computes $f$ with constant probability, say $2/3$. Given an input distribution $\mu$ to $f$, the \emph{distributional information complexity} of $f$ with respect to $\mu$ is the minimum information cost of a protocol that computes $f$ with constant probability, say $2/3$, over $\mu$. More formally, we may write:
\begin{align*}
    \R(f) := \min_{\forall x,y, \ \Pr\left[\pi(x,y) = f(x,y)\right] \geq \frac{2}{3}} \ \mathsf{CC}(\pi) \hspace{6.5mm} \text{ and } \hspace{6.5mm}  \IC_{\mu}(f) := \inf_{\underset{(x,y)\sim \mu}{\Pr}\left[\pi(x,y) = f(x,y)\right] \geq \frac{2}{3}} \ \mathsf{IC}(\pi).
\end{align*}

%% file: 3-approach.tex
\section{Approach}

Our results adopt a novel application of the \emph{embedding argument} of Barak, Braverman, Chen, and Rao \cite{BBCR13}. The first step is to reinterpret our main theorems as protocol-simulation statements: given a protocol for a composed function $g \circ f^n$, we aim to extract a protocol for a single instance of $f$ whose information cost is appropriately bounded. More formally, we can paraphrase our main theorems as follows.

\begin{theorem}[Reformulation of \Cref{thm:lifting_intro}]
Let $f:\mathcal{X}\times\mathcal{Y}\to\{0,1\}$ and let $g:\{0,1\}^n\to\mathbb{R}$. Suppose $\pi$ is a protocol that computes $g\circ f^n$ with probability $2/3$ using $C$ bits of communication. Let $\mu$ be an input distribution for $f$ with $\E_{\mu}[f(X,Y)] = \frac{1}{2}.$ Then, there exists a protocol $\eta$ that computes $f$ over $\mu$ with probability $3/5$ and has information cost at most $\frac{20}{\Inf(g)}\cdot (C+n).$
\label{thm:lifting}
\end{theorem}

\begin{theorem}[Reformulation of \Cref{thm:sym_intro}]
Let $f:\mathcal{X}\times\mathcal{Y}\to\{0,1\}$ and let $g:\{0,1\}^n \rightarrow \mathbb{R}$ be a non-constant symmetric function. Suppose $\pi$ is a protocol that computes $g\circ f^n$ with probability $2/3$ using $C$ bits of communication. Let $\mu$ be an input distribution for $f$. Then, there exists a protocol $\eta$ that computes $f$ over $\mu$ with probability $2/3$ and has information cost at most
$\frac{4C}{\switch(g)}+8.$
\label{thm:sym}
\end{theorem}

\subsection{Input Embedding Framework}
\label{subsec:BBCR}


We now recall the input embedding argument  of \cite{BBCR13}. Consider a protocol $\pi$ operating on $n$ input coordinates $(X_1,Y_1),\ldots,(X_n,Y_n)$ drawn independently from $\mu$. For each $i\in[n]$, we construct a protocol $\pi_i$ for a single input pair $(X,Y)\sim\mu$ as follows.

\begin{enumerate}
\item Use public randomness to sample $X_{<i}$ and $Y_{>i}$.
\item Alice sets $X_i=X$ and Bob sets $Y_i=Y$.
\item Alice privately draws $X_{>i}$ conditioned on $Y_{>i}$. 
\item Bob privately draws $Y_{<i}$ conditioned on $X_{<i}$.
\item Alice and Bob execute $\pi$ on inputs $\overline{X} = (X_1,\ldots,X_n)$ and $\overline{Y} = (Y_1,\ldots,Y_n)$.
\item Recover the value of $f(X,Y)$ from the output of $\pi$.
\end{enumerate}

A key feature of this construction is that the distribution of $(\overline{X},\overline{Y})$ is identical for \emph{every} $i \in [n]$, which means $\pi_i$ embeds $(X,Y)$ into coordinate $i$ while completely hiding its location. Consequently, when $\pi$ is invoked, it cannot distinguish which protocol $\pi_i$ generated the input which intuitively prompt $\pi$ to solve $f$ at all $n$ coordinates. Quantitatively, it follows that the expected information cost of $\pi_i$ drops to only a $1/n$ fraction of the information cost of $\pi$ (proof deferred to later sections).


We note that the ``recovery'' step (Line 6) is essential as the goal of $\pi_i$ is to compute $f(X,Y)$, and may require further communication. Its implementation may depend on the objective of $\pi$. For example, if $\pi$ computes the product function
$(f(X_1,Y_1),\ldots,f(X_n,Y_n))$,
then one simply reads off the $i$-th coordinate of the output. On the other hand, if $\pi$ computes the XOR function
$f(X_1,Y_1)\oplus \cdots \oplus f(X_n,Y_n)$,
then Alice can additionally compute
$f(X_1,Y_1)\oplus \cdots \oplus f(X_{i-1},Y_{i-1})$,
while Bob computes
$f(X_{i+1},Y_{i+1})\oplus \cdots \oplus f(X_n,Y_n)$.
After exchanging these values, they can recover
$f(X_i,Y_i)$.


\subsection{Proof Overview}
\label{subsec:outline}

We now give an overview of our proofs which base on simple-yet-non-standard modifications to the above embedding framework. The formal arguments are presented in \Cref{sec:lifting,sec:majority}.

\paragraph{Lifting-by-Influence.}
Let us first attempt to apply the input embedding argument as is, and see where the roadblocks are at. Consider, for instance, the protocol $\pi_1$. After the public-randomness phase, Alice learns $z_{>1}$, and after executing $\pi$, both players obtain the value of $g(z_1,\ldots,z_n)$. They then shall attempt to recover $z_1$ from this information.

However, recovery is only successful when
$g(0,z_2,\ldots,z_n)\neq g(1,z_2,\ldots,z_n)$.
Since each $z_i$ distributes independently and uniformly in $\{0,1\}$, this event occurs with probability exactly $\Inf_1(g)$. Otherwise if $g(0,z_2,\ldots,z_n)= g(1,z_2,\ldots,z_n)$, the players cannot distinguish whether $z_1=0$ or $z_1=1$; hence, may output a random bit and be correct with probability $\frac{1}{2}$. This results in an advantage of $\Inf_1(g)$ for the protocol $\pi_1$. Consequently, when $i$ is chosen uniformly at random and the protocol $\pi_i$ is run, the players can achieve advantage
$\frac{1}{n}\cdot \sum_{i=1}^n \Inf_i(g)=
\frac{\Inf(g)}{n}$.

Our goal, however, is to obtain constant advantage. A standard amplification trick resolves this issue. For each $\pi_i$, the players may repeat the embedding independently $\Theta\!\left(\frac{1}{\Inf_i(g)}\right)$ times using fresh public randomness to obtain new sets of $z_{-i}$. This boosts the probability that at least one recovery is successful to high constant, say, $0.999$. The drawback is that the information cost of $\pi_i$ now incurs an additional factor of $O\!\left(\frac{1}{\Inf_i(g)}\right)$ as the protocol $\pi$ is executed that many times.

To compensate for such blow-up, the final protocol $\eta$ shall run $\pi_i$ where $i \in [n]$ is chosen  with probability proportional to $\Inf_i(g)$. This introduces a normalization factor $\sum_{i=1}^n \Inf_i(g)=\Inf(g)$,  which ultimately results in the denominator of $\Inf(g)$ in the statement of \Cref{thm:lifting}.

\paragraph{Symmetric Composition.}
Suppose now that the gadget $g$ is symmetric, and $\switch(g) = d$ is attained via a transition at $d \leq \lfloor \frac{n+1}{2}\rfloor$ such that $\hat{g}(d) \ne \hat{g}(d-1)$. Our proof strategy relies on the observation that the embedding argument remains valid even if we embed the true input into only a \emph{subset} of coordinates rather than all. Specifically, we partition the $n$ input coordinates of $g \circ f^n$ into three disjoint parts.
\begin{enumerate}
    \item The first $d$ coordinates are the ``important'' part in which we will embed the input of $f$ into.
    \item The next $d-1$ coordinates are chosen strategically so that combining with the first $d$ coordinates yields a Hamming weight of $d-1$ or $d$.
    \item The remaining $n-2d+1$ coordinates are hardwired so that each evaluates to 0.
\end{enumerate}

More specifically, consider the following protocol $\pi'_1$ for computing $f(X,Y)$. Alice and Bob embed the true input $(X,Y) \sim \mu$ into the coordinate $(X_1,Y_1)$. Using public randomness, they sample only the next $d-1$ coordinates $(Y_2,\ldots,Y_d)$ and after which Alice privately samples $(X_2,\ldots,X_d)$.
Alice then computes $\gamma=f(X_2,Y_2)+\cdots+f(X_d,Y_d)$, sends it to Bob, and together they hardwire the next $d-1$ coordinates so that exactly $\gamma$ of them evaluate to $0$ and exactly $d-1-\gamma$ evaluate to $1$. Finally, the remaining $n-2d+1$ coordinates are hardwired so they all evaluate to 0. 

Following this construction, the Hamming weight of $(f(X_1,Y_1),\ldots,f(X_n,Y_n))$ is either $d-1$ or $d$, or more precisely $d-1+f(X,Y)$. The players then run $\pi$ which compute $g \circ f^n$. As $g$ has different evaluations at Hamming weight $d-1$ and $d$, they must be able to distinguish between $f(X,Y)=0$ and $f(X,Y)=1$.

Finally, as we embed $(X,Y)$ into one of the $d$ important coordinates, this contributes to the denominator of $d = \switch(g)$ in the statement of \Cref{thm:sym}.

\subsection{Value-to-Point Sampling}

Both of our approaches outlined above suffer from a minor pitfall: each embedded protocol incurs an undesirable hidden additive cost during the recovery phase. To illustrate, consider the protocol $\pi'_1$ from the latter construction. In order to set the middle $d-1$ coordinates in step 2, Alice must communicate the value of $\gamma=f(X_2,Y_2)+\cdots+f(X_d,Y_d)$ to Bob, which requires an additional $O(\log d)$ bits of communication.

The situation is even more problematic for the lifting-by-influence argument. Consider the protocol $\pi_i$ in which $(X,Y)$ is embedded into the coordinate $(X_i,Y_i)$. Writing $g=z_i\cdot h+r$ where $h$ and $r$ depend on $z_{-i}$ but not on $z_i$, the players must determine whether the embedded value $Z_i=f(X_i,Y_i)$ is recoverable which is equivalent to deciding whether $h(z_{-i})=0$. However, after the sampling phase, Bob knows $z_{<i}$ while Alice knows $z_{>i}$. Since the structure of $h$ is arbitrary, determining whether $h(z_{-i})=0$ may require up to additional $O(n)$ bits of communication.

To eliminate these extra costs, we introduce a small augmentation to the public sampling phase of the embedding procedure. Rather than using public randomness solely to sample $X_{<i}$ and $Y_{>i}$, we additionally sample the realized values of $Z_{-i}$ themselves. The remaining variables are then generated privately, conditioned on these sampled values.

This modification grants a useful tradeoff. Both players now know $Z_{-i}$ without exchanging any additional information, allowing them to perform the recovery step locally. Meanwhile, conditioning on $Z_{-i}$ provably only increases the information cost by $O(1)$ bits and, more importantly, still preserves the independence structure between the input coordinates. We defer the analysis to the proof sections where the sampling procedure is incorporated directly into our embedding arguments.

%% file: 4-lifting.tex
\section{Lifting by Influence}
\label{sec:lifting}

In this section we prove Theorem~\ref{thm:lifting}. We assume its setting. Let $f:\mathcal{X} \times \mathcal{Y} \rightarrow \{0,1\}$ be an inner function and $g:\{0,1\}^n\to\mathbb{R}$ be an outer gadget. Suppose $\pi$ is a protocol that computes $g\circ f^n$ with error $1/3$ using $C$ bits of communication. Fix an input distribution $\mu$ such that $\mathbb{E}_{\mu}[f(X,Y)] = \frac{1}{2}.$ Our goal is to eventually construct a new protocol $\eta$ that computes $f$ over $(X,Y) \sim \mu$ with low information cost. We first consider a set of intermediate protocols $\{\pi_i\}_{i \in [n]}$ where each $\pi_i$ operating over the input distribution $\mu$ is constructed as follows.
\\

\begin{algorithm}[H]
\caption{A protocol $\pi_i$ for computing $f(X,Y)$ over the input distribution $(X,Y) \sim \mu$}
Write $g = z_i \cdot h+r$ where $h,r$ solely depends on $z_{-i}$ \\
\For{$\frac{10}{\Inf_i(g)}$ rounds}{
Players use public randomness to draw $Z_{-i}$ i.i.d. uniformly from $\{0,1\}$ \\
They also publicly draw $X_{<i}$ conditioned on $Z_{<i}$ and $Y_{>i}$ conditioned on $Z_{>i}$ \\
Alice sets $X_i = X$. Bob sets $Y_i = Y$. Denote a random variable $Z_i = f(X_i, Y_i)$. \\
Alice privately draws $X_{>i}$ conditioned on $ Z_{>i}Y_{>i}$. \\
Bob privately draws $Y_{<i}$ conditioned on $Z_{<i}X_{<i}$. \\
Alice and Bob simulate $\pi$ on $\overline{X} = (X_1,\ldots,X_n)$ and $\overline{Y} = (Y_1,\ldots,Y_n)$ to obtain the transcript $M$. Let $b$ denote their belief of $g \circ f^n(\overline{X}, \overline{Y})$ after obtaining $M$. \\
    \If{$h(Z_{-i}) = 0$}{
        this round is a fail
    }
Otherwise, players can successfully recover $f(X,Y) = f(X_i,Y_i) = \frac{1}{h} \cdot (b-r)$.
}
\If{at least one round is a success}{
\Return the answer $f(X,Y)$ of that round
}
\Else {
\Return a random bit}
\end{algorithm}
\vspace{4mm}

\begin{remark}
In fact, the players can decide if a round fails by executing Lines 9-10 \emph{without} simulating $\pi$ in Line 8. With this modification, the information cost remains unchanged while the communication cost is reduced to $\mathsf{CC}(\pi)$ since the players need to simulate $\pi$ only once.
\end{remark}

Before proceeding to the analysis, we record the key property underlying the construction of $\pi_i$.

\begin{observation}
For every $i\in[n]$, the joint distribution of $(\overline X,\overline Y,\overline Z,M)$ generated during a round of $\pi_i$ (i.e. lines 3-11) is identical to the distribution of $(\overline X,\overline Y,\overline Z,M)$ generated by $\pi$.
\label{obs:lift_equiv_dist}
\end{observation}

\begin{proof}
For $\pi$, each triple $(X_j,Y_j,Z_j)$ is drawn independently where $(X_j,Y_j)\sim\mu$ and $Z_j=f(X_j,Y_j)$. 

Now consider a round of $\pi_i$ in lines 3-11. First, observe that each $(X_j,Y_j,Z_j)$ is drawn independently. The embedded pair $(X_i,Y_i)$ is drawn from $\mu$ which means $Z_i=f(X_i,Y_i)$ and $(X_i,Y_i,Z_i)$ have exactly the same distribution as in $\pi$. Consider next a coordinate $j<i$. The protocol first samples $Z_j$ according to its correct marginal distribution, i.e. uniform between $\{0,1\}$. It then samples $X_j \sim \mu(X\mid Z_j)$ and $Y_j \sim \mu(Y\mid X_jZ_j)$. By the chain rule of probability, the resulting triple $(X_j,Y_j,Z_j)$ has exactly the same joint distribution as under $\mu$ and hence under $\pi$. An identical argument applies to coordinates $j>i$, except that $Y_j$ is sampled before $X_j$.

As a result, the joint distribution of $(\overline X,\overline Y,\overline Z)$ is identical in a round of $\pi_i$ and in $\pi$. Since the transcript $M$ is generated by running the same protocol $\pi$ on these identically distributed inputs $(\overline X,\overline Y)$, it follows that $M$ also distributes identically.
\end{proof}

Now we argue the behaviors of $\pi_i$.

\begin{claim} \label{clm:error_lifting} The protocol $\pi_i$ computes $f(X,Y)$ with probability $0.6$.
\end{claim}

\begin{proof}
Consider one round of lines 3-11. Each $Z_{-i}$ is drawn i.i.d. uniformly from $\{0,1\}$, and this round fails iff $h(Z_{-i}) \ne 0$. Recall that $g = z_i\cdot h + r$. This means $h \ne 0$ is precisely the scenario where changing from $z_i=0$ to $z_i=1$ changes the value of $g$, which occurs with probability precisely $\Inf_i(g)$. In other words, a single round fails with probability $1-\Inf_i(g)$. As a result, the probability that all $\frac{10}{\Inf_i(g)}$ independent rounds fail is $(1-\Inf_i(g))^{\frac{10}{\Inf_i(g)}} < e^{-10} < 0.001.$

In an event that there is a successful round, $\pi$ shall compute $b = z_i \cdot h + r$ with probability $2/3$. Assuming that, the recovery step $\frac{1}{h} \cdot (b-r)$ correctly retrieves $z_i = f(X_i,Y_i) = f(X,Y)$. Overall, by the union bound, the error occurs with probability $\frac{1}{3}+0.001 < 0.4$.
\end{proof}

\begin{claim} \label{clm:IC_lifting} The protocol $\pi_i$ has information cost bounded by \begin{align*}
    \IC(\pi_i) \leq \frac{10}{\Inf_i(g)}\cdot \left[I_{\pi}(M:X_i \mid X_{<i}\,\overline{Y}\,\overline{Z}) + I_{\pi}(M:Y_i \mid Y_{>i}\,\overline{X}\,\overline{Z}) + 2\right].
\end{align*}
\end{claim}

\begin{proof}
Let us first analyze a single independent round. Consider Bob's perspective of the protocol where his input is $Y_i$, and from the public randomness he additionally observes $X_{<i}$, $Y_{>i}$, and $Z_{-i}$. Thus, his side of the information cost is $I_{\pi_i}(M:X_i \mid X_{<i} Y_{\ge i} Z_{-i}).$ We bound this quantity by
\begin{align*}
    & I_{\pi_i}(M:X_i \mid X_{<i} Y_{\ge i} Z_{-i}) \\
    & = I_{\pi}(M:X_i \mid X_{<i} Y_{\ge i} Z_{-i}) \tag{\Cref{obs:lift_equiv_dist}}\\
    &\leq I_{\pi}(MY_{<i}Z_i:X_i \mid X_{<i} Y_{\ge i} Z_{-i}) \\
    &=
    I_{\pi}(Y_{<i}:X_i \mid X_{<i} Y_{\ge i} Z_{-i}) +
    I_{\pi}(Z_i:X_i \mid X_{<i} \overline{Y} Z_{-i}) +
    I_{\pi}(M:X_i \mid X_{<i} \overline{Y}\,\overline{Z}) \tag{chain rule}
\end{align*}
The first term vanishes as 
$I_{\pi}(Y_{<i}:X_i \mid X_{<i} Y_{\ge i} Z_{-i}) = 0.$ To see this, observe that $Y_{<i} \perp X_i \mid X_{<i} Y_{\ge i}$ as each coordinate is drawn i.i.d. from $\mu$. Conditioning further on $Z_{-i}$ does not change the independence because each $Z_j$ only depends on $(X_j,Y_j)$ but not other coordinates.
For the second term, since $Z_i$ is a single bit, we can bound
$I_{\pi}(Z_i:X_i \mid X_{<i} \overline{Y} Z_{-i}) \leq |Z_i|=1.$
Hence, we obtain
\begin{align*}
I_{\pi}(M:X_i \mid X_{<i} Y_{\ge i} Z_{-i})
\leq
I_{\pi}(M:X_i \mid X_{<i} \overline{Y}\,\overline{Z}) + 1.
\end{align*}
Applying the symmetric argument for Alice's side, the information cost of a round is bounded by
\begin{align*}
I_{\pi}(M:X_i \mid X_{<i}\,\overline{Y}\,\overline{Z}) + I_{\pi}(M:Y_i \mid Y_{>i}\,\overline{X}\,\overline{Z}) + 2.
\end{align*}Since the $\pi_i$ performs
$\frac{10}{\Inf_i(g)}$
independent runs, the desired bound follows.
\end{proof}

We now finally prove \Cref{thm:lifting}.

\begin{proof}[Proof of \Cref{thm:lifting}] We construct the protocol $\eta$ which proceeds by drawing an index $i \in [n]$ with probability $\frac{\Inf_i(g)}{\Inf(g)}$, and running $\pi_i$. Using \Cref{clm:error_lifting}, $\eta$ succeeds with probability $0.6$. The information cost of $\eta$ can be bounded via
\begin{align*}
\IC(\eta) & = \sum_{i=1}^{n} \frac{\Inf_i(g)}{\Inf(g)} \cdot \IC(\pi_i) \\ 
& \leq \frac{1}{\Inf(g)} \cdot \sum_{i=1}^{n} \left(I_{\pi}(M:X_i \mid X_{<i} \overline{Y} \, \overline{Z}) + I_{\pi}(M:Y_i \mid Y_{>i} \overline{X} \, \overline{Z}) + 2\right) \tag{\Cref{clm:IC_lifting}}\\
& = \frac{10}{\Inf(g)} \cdot \left[I_{\pi}(M:\overline{X} \mid \overline \, \overline{Z}) + I_{\pi}(M:\overline{Y} \mid \overline{X} \,\overline{Z}) + 2n\right] \tag{chain rule} \\
& \leq \frac{20\cdot (C+n)}{\Inf(g)}
\end{align*}
where the final inequality follows $I_{\pi}(M:\overline{X} \mid \overline{Y} \, \overline{Z}), I_{\pi}(M:\overline{Y} \mid \overline{X} \,\overline{Z}) \leq |M| \leq C$.
\end{proof}

%% file: 5-symmetric.tex
\section{Symmetric Composition}
\label{sec:majority}

In this section we prove \Cref{thm:sym}. Suppose now that the gadget $g$ is symmetric and $\switch(g) = d$ is attained via $d \in [n]$ such that $\hat{g}(d) \ne \hat{g}(d-1)$. We may assume wlog that $d \leq \lfloor \frac{n+1}{2}\rfloor$; otherwise we can consider $\overline{g}(z_1,\ldots,z_n) := g(1-z_1,\ldots,1-z_n)$ instead. Let $\pi$ be a protocol that computes $g \circ f^n$ with error $1/3$ using $C$ bits of communication. Fix an input distribution $\mu$ to $f$. Our goal is to eventually construct a new protocol $\eta$ that computes $f$ over $(X,Y) \sim \mu$. 

Let $0_A,1_A \in \mathcal{X}$ and $0_B, 1_B \in \mathcal{Y}$ denote special inputs to Alice and Bob such that $f(0_A,0_B) = 0$ and $f(1_A,1_B) = 1$. These inputs must exist as otherwise $f$ would have been a constant, to which no communication is required for computing $f$ nor $g \circ f^n$. We now construct a hard input distribution $\theta := \theta(\mu)$ to $g\circ f^n$, and re-interpret $\pi$ as a distribution when executed on $(\overline{X},\overline{Y}) \sim \theta$.

\begin{figure}[H]
    \centering
    \begin{tabular}{|p{15.5cm}|}
    \hline ~\\
    \multicolumn{1}{|c|}{\textbf{An input distribution $\theta$ and the protocol $\pi$ for computing $g \circ f^n$ over $\theta$.}} \\ 
    \\ \hline
    \vspace{-2.85mm}
    \begin{enumerate}
    \item For each $i \in [d]$, draw $(X_i,Y_i) \sim \mu$ i.i.d.
    \item Draw an independent bit $b \in \{0,1\}$ u.a.r. and denote $\tau = \sum_{i \in [d]} f(X_i,Y_i) - b$
    \item Alice sets $\overline{X} = (X_1,\ldots,X_d,\underbrace{0_A,\ldots,0_A}_{\tau},\underbrace{1_A,\ldots,1_A}_{d-1-\tau}, \underbrace{0_A,\ldots,0_A}_{n+1-2d})$
    \item Bob sets $\overline{Y} = (Y_1,\ldots,Y_d,\underbrace{0_B,\ldots,0_B}_{\tau},\underbrace{1_B,\ldots,1_B}_{d-1-\tau}, \underbrace{0_B,\ldots,0_B}_{n+1-2d})$
    \item Let $\pi(\cdot)$ denote the probability space when applying the protocol $\pi$ over $(\overline{X},\overline{Y}) \sim \theta$.
\end{enumerate}

    \vspace{1mm}
    \\
    \hline
    \end{tabular}
    \caption{A hard distribution $\theta$ for $g\circ f^n$ and a distributional view of $\pi$.}
    \label{fig:theta}
\end{figure}

\begin{remark} 
The sampling procedure may fail to produce $(\overline{X},\overline{Y})$ if $\tau=d$. This turns out not to be a real issue as this event occurs with probability only $2^{-(d+1)}$. Moreover, in our actual analysis, we shall restrict the underlying probability space so that this failure event never occurs.
\end{remark}

Let us first provide the intuition behind the distribution $\theta$, inspired by the protocol $\pi'_1$ outlined in \Cref{subsec:outline}. Under $\theta$, the Hamming weight of $f^n(\overline{X},\overline{Y})$ is $d-1+b$ which is dictated by the hidden bit $b \in \{0,1\}$. As the evaluation of $g$ at Hamming weights $d-1$ and $d$ differ, an effective protocol for $g \circ f^n$ must be able to determine $b$. While the players may fully observe the hardwired coordinates prior to executing any communication, they only learn the value of $\tau=\sum_{i\in[d]} f(X_i,Y_i)-b$. Thus, their task remains to decide whether $\sum_{i\in[d]} f(X_i,Y_i)$ is $\tau$ or $\tau+1$. Since each of the first $d$ coordinates are drawn independently and the additive gap of one is vanishingly small, any successful protocol may intuitively need to recover all values $f(X_1,Y_1),\ldots,f(X_d,Y_d)$. This makes the task resemble the direct-sum problem and thereby should require $d$ time the information of computing $f$.

The remainder of the proof formalizes this intuition. Similar to before, we first write down an intermediate protocol $\pi_i$ for each $i \in [d]$.
\\

\begin{algorithm}[H]
\caption{A protocol $\pi_i$ for computing $f(X,Y)$ over the input distribution $(X,Y) \sim \mu$}
Players use public randomness to draw $Z_{[d]\setminus \{i\}}$ i.i.d. uniformly from $\{0,1\}$ \\
They also publicly draw $X_{<i}$ conditioned on $Z_{<i}$ and $Y_{[i+1,d]}$ conditioned on $Z_{[i+1,d]}$ \\
Alice embeds $X_i = X$. Bob embeds $Y_i = Y$. Denote a random variable $Z_i = f(X_i, Y_i)$ \\
Alice privately draws $X_{[i+1,d]}$ conditioned on $Z_{[i+1,d]}Y_{[i+1,d]}$ \\
Bob privately draws $Y_{<i}$ conditioned on $Z_{<i}X_{<i}$ \\
Both players locally compute $\gamma = \sum_{j \in [d]\setminus \{i\}} Z_j$ \\
Alice sets $\overline{X} = (X_1,\ldots,X_d,\underbrace{0_A,\ldots,0_A}_{\gamma},\underbrace{1_A,\ldots,1_A}_{d-1-\gamma}, \underbrace{0_A,\ldots,0_A}_{n+1-2d})$ \\
Bob sets $\overline{Y} = (Y_1,\ldots,Y_d,\underbrace{0_B,\ldots,0_B}_{\gamma},\underbrace{1_B,\ldots,1_B}_{d-1-\gamma}, \underbrace{0_B,\ldots,0_B}_{n+1-2d})$ \\
Alice and Bob simulate $\pi$ over the inputs $(\overline{X},\overline{Y})$ to obtain a transcript $M$. Let $\ell$ denote their belief of $g \circ f^n(\overline{X},\overline{Y})$ after obtaining the transcript. \\
\If{$\hat{g}(d-1) = \ell$}{
\Return 0}
\ElseIf {$\hat{g}(d) = \ell$}{
\Return 1
}
\Else{\Return fail}
\end{algorithm}

\vspace{4mm}


We emphasize that unlike in \Cref{obs:lift_equiv_dist}, the distributions $\pi_i(\overline X,\overline Y)$ here are \emph{not} identical across different choices of $i$. This symmetry was crucial in the information cost analysis of \Cref{clm:IC_lifting} as it allowed every $\pi_i$ to be replaced by the common distribution $\pi$. To overcome this obstacle, we instead show that each $\pi_i$ here corresponds to exactly a \emph{half} of the probability space induced by $\theta$ and $\pi$ (recall \Cref{fig:theta}), which turns out to be sufficient for the remaining of our proofs.

\begin{observation}
For every $i\in[d]$, there exists an event $W_i$ which occurs with probability $\frac{1}{2}$ and $\pi_i(\overline X,\overline Y,Z_{\leq d},M)=(\pi\mid W_i)(\overline X,\overline Y,Z_{\leq d},M).$
\label{obs:dist_equiv_maj}
\end{observation}

\begin{proof} Observe that the first $d$ coordinates of $\pi_i$ are drawn independently according to $\mu$. Therefore, we have $\pi_i(X_{\leq d},Y_{\leq d}) = \mu^d = \theta(X_{\leq d},Y_{\leq d})$ and $Z_{\leq d}$ has the same distribution under both $\pi_i$ and $\theta$. 

Recall that under the distribution $\theta$, the bit $b \in \{0,1\}$ is uniformly random and is independent of $(X_{\leq d},Y_{\leq d})$, and therefore is independent of $Z_{\leq d}$. Let $W_i$ be the event that $b = Z_i$ which occurs with probability $\frac{1}{2}$. 

Conditioned on $W_i$ and the same realization of $Z_{\leq d}$, the quantities $\tau$ (in $\theta$) and $\gamma$ (in $\pi_i$) coincide as both are equal to $\sum_{j \in [d]\setminus \{i\}} Z_j$. Consequently, the next $d-1$ coordinates are generated under the same distribution under both $\pi_i$ and $\theta \mid W_i$. At last, the final $n+1-2d$ coordinates are identical as they are hardwired $0$-inputs. Together, it follows that $\pi_i(\overline X,\overline Y)
= (\theta\mid W_i)(\overline X,\overline Y).$

Since the transcript $M$ is generated by running the same protocol $\pi$ on these identically distributed inputs $(\overline X,\overline Y)$, it follows that $M$ also distributes identically in $\pi_i$ and $\pi\mid W_i$.
\end{proof}

\begin{claim} \label{clm:main_maj} The protocol $\pi_i$ computes $f(X,Y)$ with probability $2/3$,
and has information cost bounded by \begin{align*}
    \IC(\pi_i) \leq 2 \cdot I_{\pi}(M:X_i \mid X_{<i}\,Y_{\leq d}\,Z_{\leq d}) + 2 \cdot I_{\pi}(M:Y_i \mid Y_{[i+1,d]}\,X_{\leq d}\,Z_{\leq d}) + 8.
\end{align*}
\end{claim}

\begin{proof} Recall that $\pi$ is correct with probability $2/3$ for computing $g \circ f^n(\overline X, \overline Y)$. Assume that this is the case. Observe that according to $\pi_i$, the Hamming weight of $f^n(\overline{X},\overline{Y})$ is $d-1+f(X,Y) \in \{d-1,d\}$. As $g$ has different evaluations at Hamming weights $d-1$ and $d$, i.e., $\hat{g}(d-1) \ne \hat{g}(d)$, the value $\ell$ shall let the players faithfully recover $f(X,Y)$.

We now bound the information cost of $\pi_i$. Consider Bob's perspective of the protocol where his input is $Y_i$, and from the public randomness he additionally observes $X_{<i}$, $Y_{[i+1,d]}$, and $Z_{[d]\setminus \{i\}}$. Therefore, his side of the information cost is $I_{\pi_i}(M:X_i \mid X_{<i} Y_{[i+1,d]} Z_{[d]\setminus \{i\}})$. Using \Cref{obs:dist_equiv_maj} and \Cref{lem:helper}, we can derive 
\begin{align}
    I_{\pi_i}(M:X_i \mid X_{<i} Y_{[i+1,d]} Z_{[d]\setminus \{i\}}) & = I_{\pi \mid W_i}(M:X_i \mid X_{<i} Y_{[i+1,d]} Z_{[d]\setminus \{i\}})  \nonumber \\
    & \leq 2 \cdot I_{\pi}(M:X_i \mid X_{<i} Y_{[i+1,d]} Z_{[d]\setminus \{i\}}) + 2. \label{eq1}
\end{align}
Following a similar derivation as in \Cref{clm:IC_lifting} (with $n$ replaced by $d$), we shall have $I_{\pi}(M:X_i \mid X_{<i} Y_{[i+1,d]} Z_{[d]\setminus \{i\}}) \leq I_{\pi}(M:X_i \mid X_{<i}\,Y_{\leq d}\,Z_{\leq d}) + 1$. Combining this with (\ref{eq1}) and the symmetric term for Alice's side, the information cost of $\pi_i$ is bounded by
\begin{align*}
 \IC(\pi_i) \leq 2 \cdot I_{\pi}(M:X_i \mid X_{<i}\,Y_{\leq d}\,Z_{\leq d}) + 2 \cdot I_{\pi}(M:Y_i \mid Y_{[i+1,d]}\,X_{\leq d}\,Z_{\leq d}) + 8
\end{align*}
as wished.
\end{proof}

 We now finally prove \Cref{thm:sym}.

\begin{proof}[Proof of \Cref{thm:sym}] Consider the protocol $\eta$ which proceeds by running $\pi_i$ for a uniformly random $i \in [d]$. Using \Cref{clm:main_maj}, the protocol $\eta$ succeeds with probability $2/3$ and has information cost bounded by
\begin{align*}
\IC(\eta) & = \frac{1}{d} \cdot \sum_{i=1}^{n} \IC(\pi_i) \\ & \leq \frac{1}{d} \cdot \sum_{i=1}^{d} \left[2 \cdot I_{\pi}(M:X_i \mid X_{<i}\,Y_{\leq d}\,Z_{\leq d}) + 2 \cdot I_{\pi}(M:Y_i \mid Y_{[i+1,d]}\,X_{\leq d}\,Z_{\leq d}) + 8 \right] \tag{\Cref{clm:main_maj}}\\
& = \frac{1}{d} \cdot \left[2 \cdot I_{\pi}(M:X_{\leq d} \mid Y_{\leq d}\,Z_{\leq d}) + 2 \cdot I_{\pi}(M:Y_{\leq d} \mid X_{\leq d}\,Z_{\leq d}) + 8d \right] \tag{chain rule} \\
& \leq \frac{4C}{d} + 8
\end{align*}
where the final inequality follows $I_{\pi}(M:X_{\leq d} \mid Y_{\leq d}\,Z_{\leq d}), I_{\pi}(M:Y_{\leq d} \mid X_{\leq d}\,Z_{\leq d}) \leq |M| \leq C$.
\end{proof}



%% file: A-even_majority.tex
\section{Majority Lemma for Even \texorpdfstring{$n$}{n}}

\label{appendix:even_maj}


We now recover the majority lemma when $n$ is even via  a simple reduction from the case when $n$ is odd. We do caseworks as follows.

\begin{itemize}
    \item $n \equiv 2 \pmod 4$. Write $n=4m-2$ for some positive integer $m$. We reduce majority on $4m-2$ coordinates to majority on $2m-1$ coordinates, which is an odd number. Given inputs $(X_1,\ldots,X_{2m-1})$ to Alice and $(Y_1,\ldots,Y_{2m-1})$ to Bob, they duplicate every coordinate and write their new inputs as $(X_1,X_1,\ldots,X_{2m-1},X_{2m-1})$ and $(Y_1,Y_1,\ldots,Y_{2m-1},Y_{2m-1})$ respectively. The new  instance has $4m-2=n$ coordinates while preserving the majority value. Since the underlying instance has an odd number of distinct coordinates, we can invoke the odd-case majority lemma to obtain the desired $\Omega(2m-1)=\Omega(n)$ scaling lower bound.

    \item $n \equiv 0 \pmod 4$. Write $n=4m$ for some positive integer $m$. We reduce this case to the previous case of $4m-2$ by hardwiring the final two coordinates to $(0_A,1_A)$ for Alice and $(0_B,1_B)$ for Bob. This does not affect the majority answer. The remaining $4m-2$ coordinates are precisely the setting handled above, giving the desired $\Omega(n)$ lower bound.
\end{itemize}